\documentclass[a4paper,11pt]{article}
\pdfoutput=1 

\usepackage{jinstpub} 

\usepackage[section]{placeins}
\RequirePackage{graphicx}

\title{Performance of a Radial Time Projection Chamber with Electroluminescence in Liquid Xenon}

\author[1]{Yuehuan Wei%
\note{Now at Sino-French Institute of Nuclear Engineering and Technology of Sun Yat-Sen University}}
\author{, Jianyang Qi, }
\author{Evan Shockley, }
\author{Haiwen Xu, }
\author[2]{Kaixuan~Ni\note{Corresponding Author}}

\affiliation{Department of Physics, University of California San Diego, La Jolla, CA, 92093, USA}

\emailAdd{nikx@physics.ucsd.edu}

\abstract{The dual-phase xenon time projection chamber (TPC) is a leading detector technology in rare event searches for dark matter and neutrino physics. The success of this type of detector technology relies on its capability to detect both primary scintillation and ionization signals from particle interactions in liquid xenon (LXe). The ionization electrons are converted into electroluminescence in the gas xenon (GXe), where a single electron can be amplified by more than 100 times in number of photons in a strong electric field. Maintaining a strong and uniform electric field in the small gas gap in large diameter TPCs is challenging. One alternative solution is to produce the electroluminescence in the LXe directly to overcome the gas gap uniformity problem. Here we report on the design and performance of a single-phase Radial TPC (RTPC) which can create and detect the electroluminescence directly in LXe. It simplifies the design and operation of the LXe TPC by using a single wire in the axial center to create the strong electric field. We present the performance of such an RTPC and discuss its limitations for potential applications. }


\arxivnumber{2111.09112} 

\begin{document}
\maketitle
\flushbottom
\section{Introduction}
\label{sec:intro}

Liquid xenon (LXe) is widely used as the detection medium for fundamental particle physics and astrophysics~\cite{Aprile:2009dv}, and is at the forefront of the rare event searches for dark matter~\cite{Cui:2017nnn,Aprile:2018dbl,XENON:2020kmp,LZ:2021xov} and neutrino physics~\cite{nEXO:2021ujk}. LXe also has potential practical applications in medical imaging~\cite{Amaudruz:2009vs, GallegoManzano:2015hkg} and nuclear safeguards~\cite{Akimov:2019ogx, Ni:2021mwa}. The detection of both ionization and scintillation signals from a radiation event interacting in LXe not only provides event discrimination but also improves the energy and position resolutions. Conventionally these two signals are either detected using separate photon and charge readout in a single-phase LXe time projection chamber (TPC), or using the two-phase xenon TPC which converts the ionization electrons into electroluminescence light in the gas xenon (GXe), subsequently detected by the same photo-sensors for the primary scintillation light. The two-phase TPC simplifies the readout and lowers the energy threshold compared with the single-phase TPC with charge readout, but it gradually becomes more and more challenging to be built with very large diameter electrodes and requires sub-mm flatness of the gas gap. 

Creating and detecting the electroluminescence directly in LXe, first demonstrated in the 1970's ~\cite{MIYAJIMA1979239, Masuda:1978tjp}, eliminate the need of a gas gap of the two-phase TPCs thus simplify the detector design, which was suggested  for future dark matter direct detection and demonstrated in small prototypes~\cite{Ye:2014gga, Aprile:2014ila}.  However, a large scale single-phase TPC using the electroluminescence in LXe hasn't been realized so far due to its delicate design using ultra-thin wires on planar electrodes, as discussed recently in ~\cite{Juyal:2019gch}. Here we present the performance of a novel design of a single-phase LXe TPC, the Radial Time Projection Chamber (RTPC), proposed in~\cite{Lin:2021izy} with simulation studies of the electroluminescence in LXe. The RTPC further simplifies the design of a LXe TPC by eliminating the need of large diameter planar electrodes and replacing them with a single wire in the axial center that creates a strong field to produce electroluminescence in LXe. 
The design and operation of the single-phase RTPC is described in Sec.\,\ref{sec:tpc}. The detector performance is summarized in Sec.\,\ref{sec:performance}, including position sensitivity, energy response, and its capability to detect low energy electronic recoils. The limitation and potential applications of this type of TPC are discussed in Sec.\,\ref{sec:potential}.

\section{Radial Time Projection Chamber (RTPC)}
\label{sec:tpc}

\subsection{RTPC Design and Field Modeling}
\label{subsec:design}

\begin{figure}
\centering
\includegraphics[width=.45\textwidth]{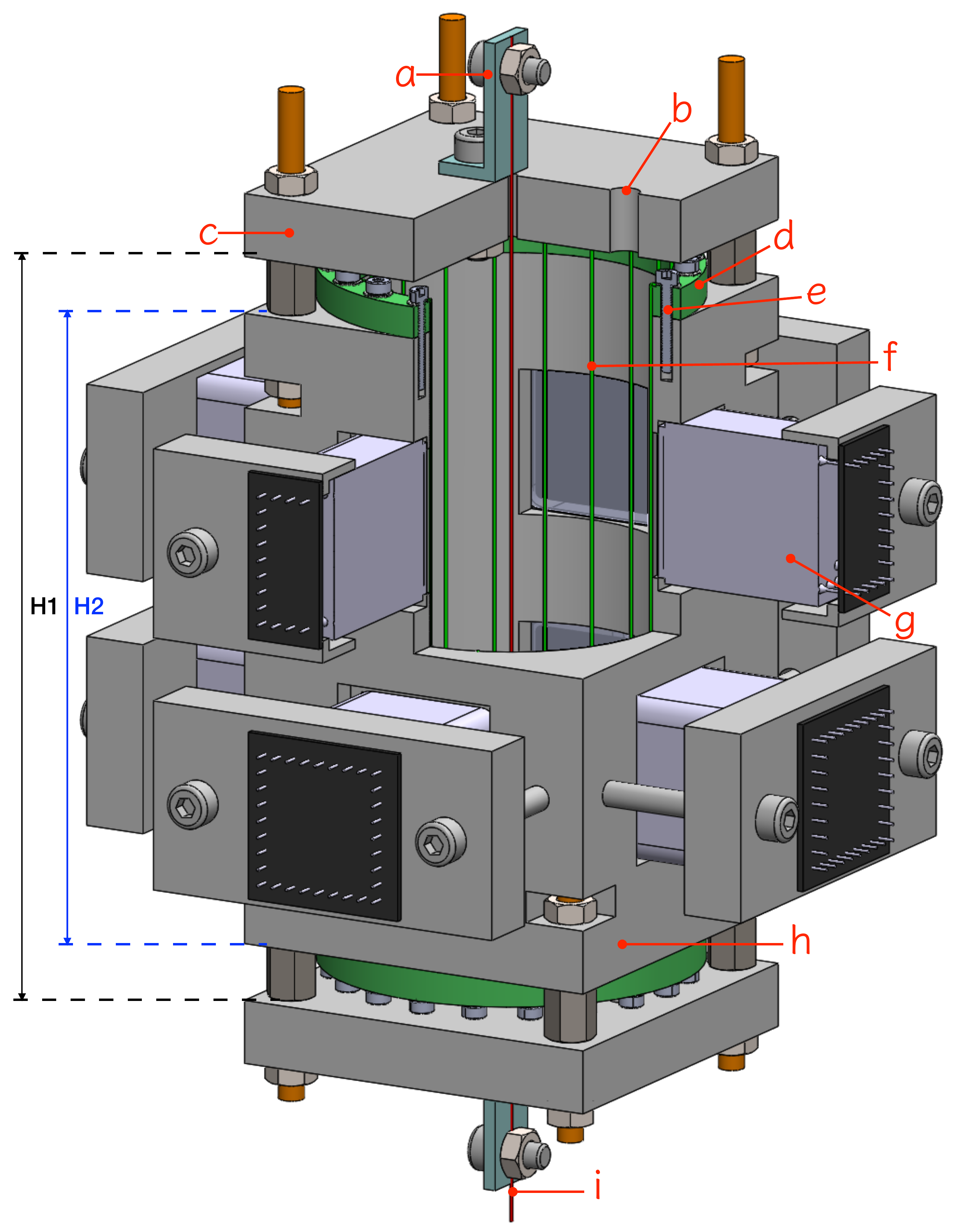}
\hspace*{.2in}
\includegraphics[width =.35\textwidth]{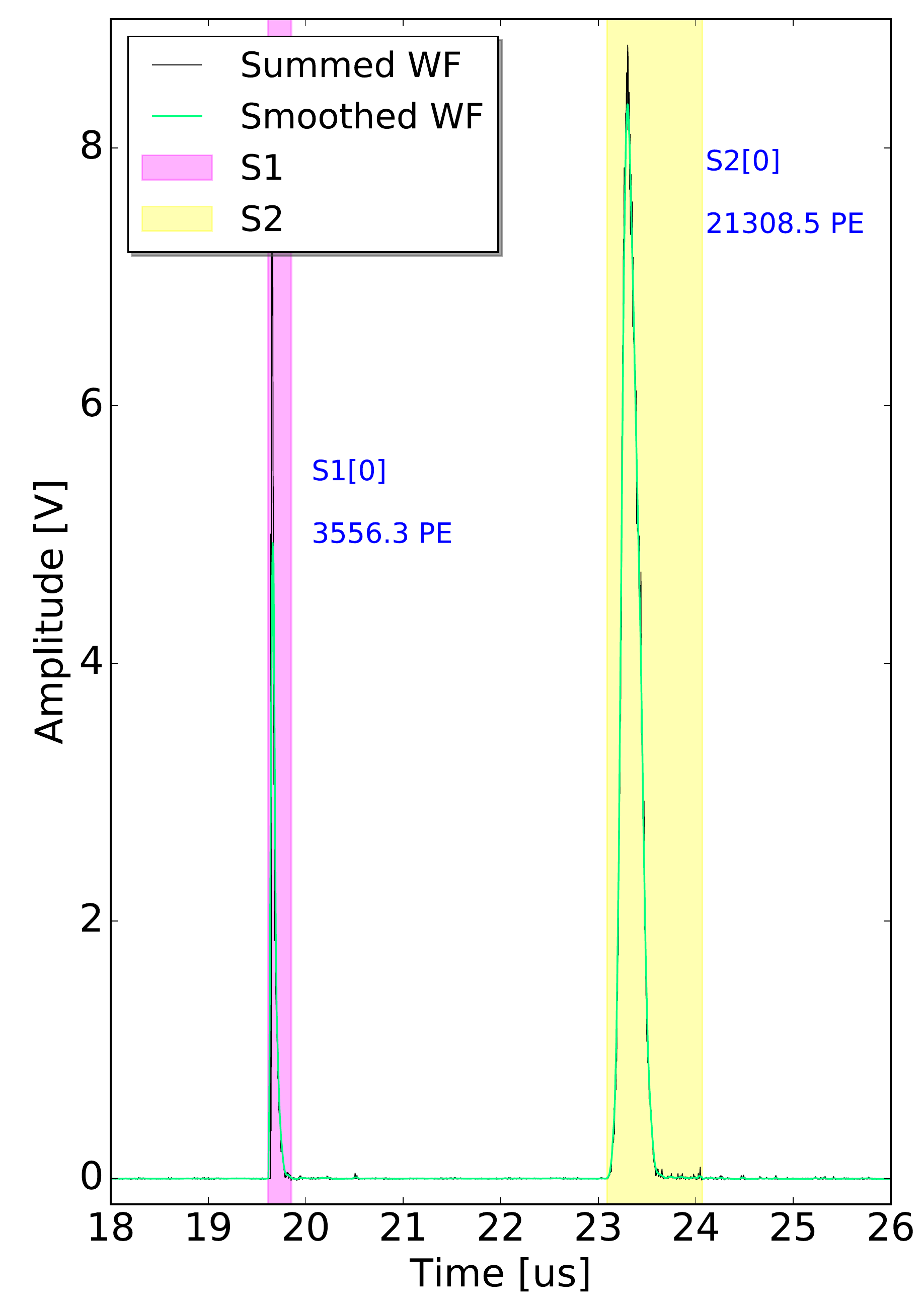}
\caption{Left: Design of the RTPC used in this study. 
The active LXe is contained within a polytetrafluoroethylene (PTFE) cylinder and viewed by eight Hamamatsu R8520 photomultiplier-tubes (PMTs). The diameter of the cylinder is 50\,mm, the maximum height (H1) is 11.8\,cm and inner height (H2) is 10\,cm. From top to bottom:
$\textit{a}$- central wire fixing plate. 
$\textit{b}$- a hole for the LED for PMT gain calibration.
$\textit{c}$- top/bottom end plates made of PTFE.
$\textit{d}$- a 5\,mm thick S.S electrode frame for fixing cathode wires. 
$\textit{e}$- screws for fixing the wires on S.S electrode frame. 
$\textit{f}$- cathode and PMT screening wires (x20).
$\textit{g}$- R8520 PMTs (x8).
$\textit{h}$- a PTFE block to define the TPC target. 
$\textit{i}$- a 25-$\mu$m diameter gold-plated tungsten wire as the anode. Right: A typical waveform from $^{137}$Cs 662\,keV gamma line with anode at +4\,kV and cathode at -750\,V.}

\label{fig:rtpc-design}
\end{figure}

The design of the single-phase RTPC used in this study is shown in Figure\,\ref{fig:rtpc-design}. 
The active LXe is defined by a 50\,mm $\times$ 100\,mm (diameter $\times$ height) cylinder made from a polytetrafluoroethylene (PTFE), corresponding to a $\sim$0.6\,kg of LXe once fully filled. The active volume is viewed by eight Hamamatsu R8520 photomultiplier-tubes (PMTs). A 25-$\mu$m diameter gold-plated tungsten wire is installed in the axial center to serve as the anode. Twenty 218-$\mu$m diameter stainless steel (S.S) wires are installed around the perimeter of the target to serve as the cathode and high voltage screening from the PMTs. 
Two 10\,mm thick PTFE plates are placed on the two ends of the cylinder with a distance of 11.8\,cm. 
High voltage is applied to the central wire from the connector on the lower end, producing a strong electric field near the anode wire for creating electroluminescence in LXe. 
The working principle of the RTPC is similar to that of a dual-phase TPC which has been introduced in \cite{Lin:2021izy}. A particle interaction in the sensitive volume produces a primary scintillation light which is referred to as the S1, the
ionized electrons are drifted to the central anode wire, and produces electroluminescence near the wire, which is called S2.

\begin{figure}[htbp]
\centering 
\includegraphics[width=.45\textwidth]{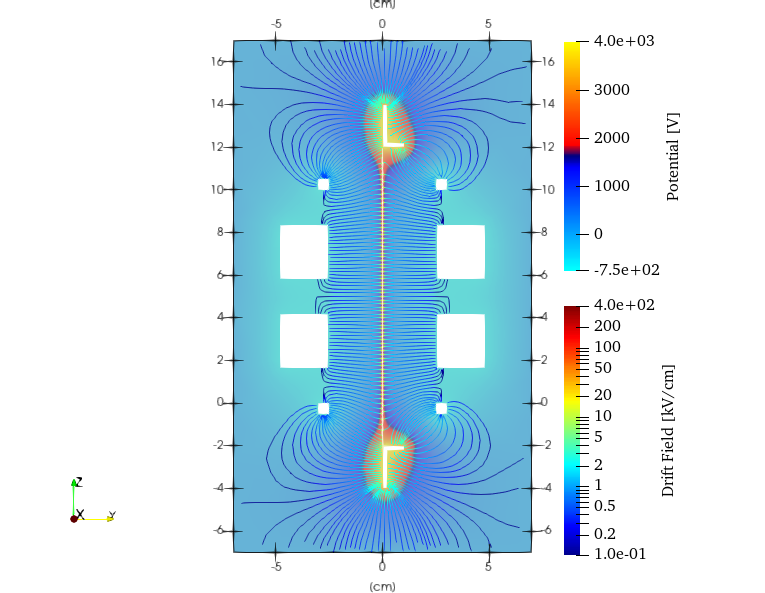}
\includegraphics[width=.53\textwidth]{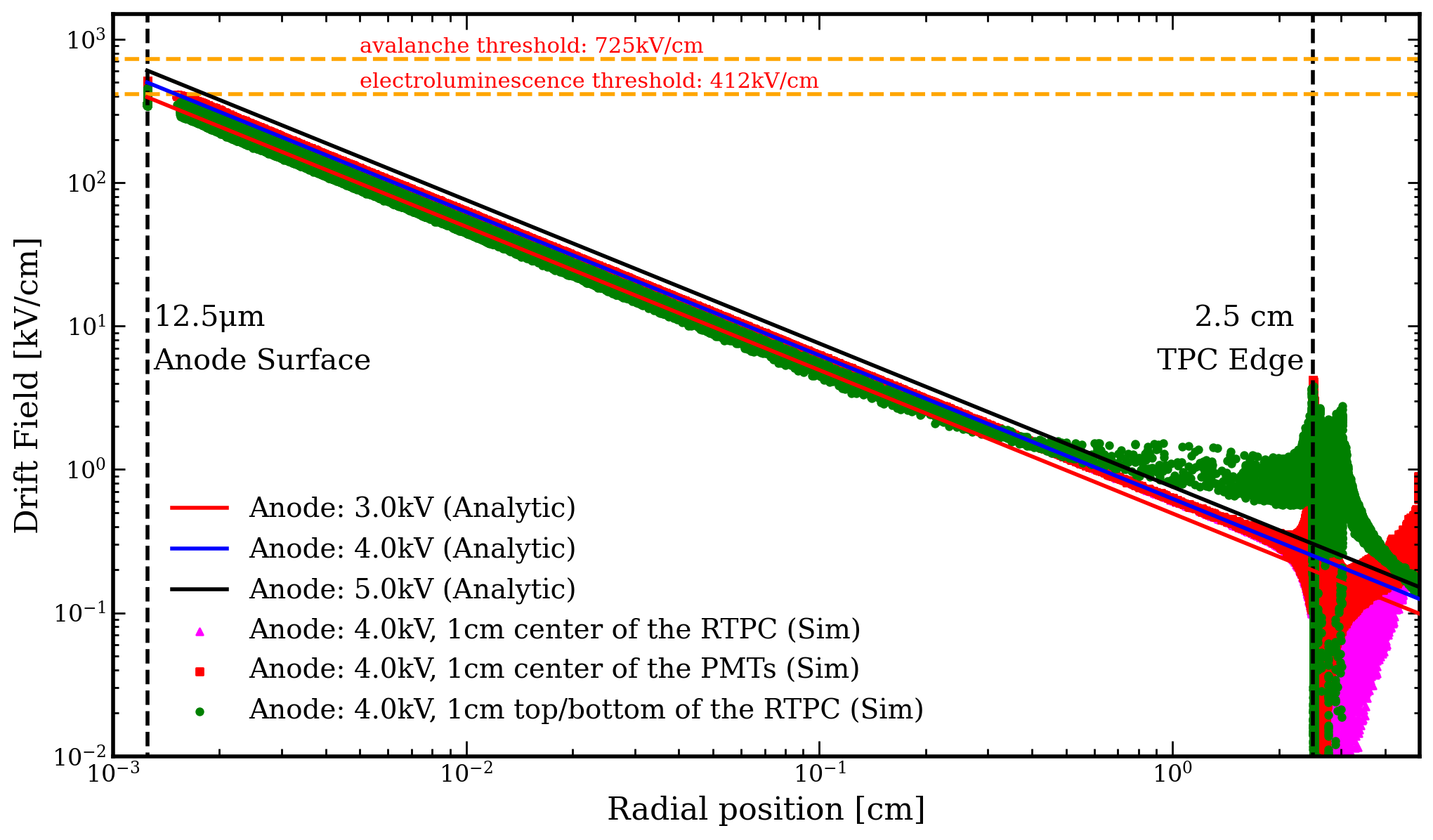}
\caption{Left: Simulated field lines using COMSOL, and the visualization is done by Paraview. Right: The analytic field as a function of RTPC radius, with anode at +3.0\,kV, +4.0\,kV and +5.0kV, cathode at -750\,V. The COMSOL simulation results are overlaid, for a 1\,cm height volume centered at the center of RTPC, PMTs, and top/bottom part of the RTPC, for a +4.0\,kV anode.}
\label{fig:rtpc-field}
\end{figure}

To realize the electroluminescence in LXe, a stronger electric field is needed compared with the requirement in GXe. The thresholds of the field strength for electroluminescence and electron avalanche were experimentally studied in \cite{Aprile:2014ila}. Given the symmetric geometry, the electric field along the radius of the RTPC can be easily calculated assuming an infinite-long tube, and is shown in the right of Figure\,\ref{fig:rtpc-field}. Based on the calculation, a +4\,kV HV on the anode wire can provide the electric field exceeding the electroluminescence threshold ($\sim 400$~kV/cm) in LXe with the cathode set at -750\,V. 
A more accurate modeling of the electric field distribution in the RTPC was performed in a 3D electric simulation taken into the actual geometry of the TPC using COMSOL. The results are overlaid in the right of Figure~\ref{fig:rtpc-field} for comparison. In fact, the simulated drift fields in most parts of the sensitive volume are quite consistent with the simplified analytical calculation. The field uniformity can be indicated by the uniformly distributed field lines as shown in the left of Figure~\ref{fig:rtpc-field}, especially for the vertical center of the RTPC.

\subsection{RTPC Operation}
\label{subsec:operation}

The RTPC is installed in the SanDiX (San Diego Xenon Detector Test System), previously used to study a hermetically-sealed LXe TPC~\cite{Wei:2020cwl} to improve the xenon purification efficiency, which has an inner S.S vessel with an inner diameter of 15\,cm to host the LXe. The anode wire and the cathode wires (PMT screening) are connected to feed-throughs from the bottom flanges of the vessel. A total of 7.7\,kg of xenon is filled into the vessel to ensure the entire TPC is covered by LXe. The xenon is constantly circulated through a SAES purifier at 4-SLPM (standard liter per minute) to improve the xenon purity over time. 

The cathode wires are set at -750\,V, close to the voltages on all eight PMTs. The PMT voltages are tuned to have the same single photo-electron (PE) gain of $\sim$\,$10^6$.  The anode wire then is biased to positive voltage to provide both the drift and the amplification field of the RTPC. During detector running, the voltage on anode wire is set as +4.0\,kV, +4.5\,kV and +5.0\,kV, respectively, for studying the property of the RTPC at different drift field and single electron gain. 

To characterise the detector performance, a $^{137}$Cs gamma ray source was deployed outside the detector, with the same height as the center of the RTPC. The raw data was digitized by CEAN V1720 FADC with a sampling frequency of 250\,MS/s. A 2-PMT trigger coincidence was required and the trigger threshold was set at $\sim$0.5\,PE.

Benefiting from the strong electric field around the central thin wire of the RTPC, a +4\,kV HV on the anode can produce a sizeable electroluminescence\,(S2) in LXe. Unlike a dual-phase TPC where the S2 is produced along a few mm field line in GXe, the S2 region is just order of a few $\mu$m near the anode wire for the RTPC. With such a short S2 region and low diffusion under a large drift field, the S2 pulses are much narrower than the dual-phase TPC. With most S2s having a width below 1\,$\mu$s at 10\% pulse height of the waveform from the RTPC. One typical waveform from $^{137}$Cs 662-keV gamma ray event depositing in the RTPC is shown in Figure\,\ref{fig:rtpc-design} (right). 

\section{Performance of the Liquid Xenon RTPC}
\label{sec:performance}
\subsection{Position Sensitivity}
\label{subsec:positions}

The RTPC detector geometry effectively swaps the two-phase TPC's $z$ coordinate with the radial coordinate $r$. This means that, in an ideal detector, $r$ can be reconstructed from the drift time, and $z$ can be inferred from the PMT hitpattern. However, since there are only 8 PMTs in this detector, the $z$ position cannot be reliably reconstructed as the hitpattern will be too coarsely binned. Nonetheless, events higher in the detector will have an S2 signal which is seen more in the top 4 PMTs than the bottom 4. Thus, we can define the S2-asymmetry parameter as the difference between the S2 light seen by the top 4 PMTs and the S2 light seen by the bottom 4 PMTs divided by the total S2 light.

\begin{equation}
    S2_{asym} = \frac{S2_{\text{4 top PMTs}}-S2_{\text{4 bot PMTs}}}{S2_{\text{4 top PMTs}}+S2_{\text{4 bot PMTs}}}
\end{equation}

\begin{figure}[htbp]
    \centering
    \includegraphics[width = 0.45\textwidth]{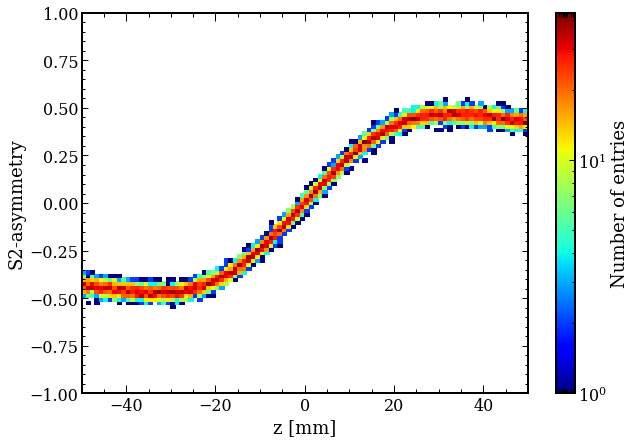}
    \includegraphics[width = 0.45\textwidth]{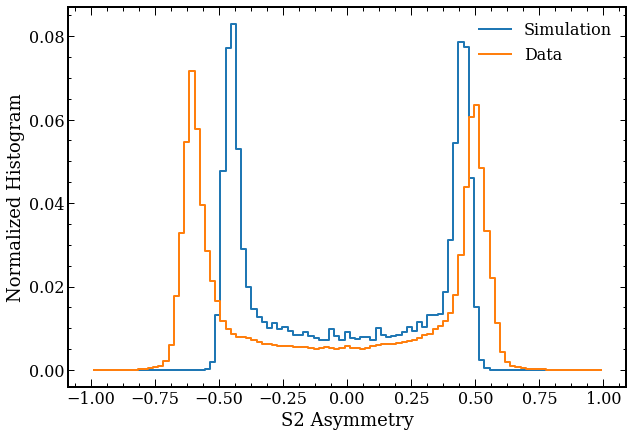}
    \caption{Left: The correlation between the S2 asymmetry parameter and $z$ from a Chroma optical simulation. Right: Asymmetry distribution from $^{137}$Cs data and Chroma simulation. }
    \label{fig:chroma}
\end{figure}

This parameter is correlated with $z$ and we can use this information to select for events away from the top and bottom PTFE plates of the RTPC. To see the correlation of the S2-asymmetry parameter with $z$, an optical simulation using the Chroma simulation package \cite{Chroma:} was performed. The simulation result is shown in the left of Figure \ref{fig:chroma}. The S2 asymmetry can represent $z$ quite well for the events generated close to the axial center of the TPC as expected, but its capability seriously deteriorates for the events with $|z|$\,>\,20\,mm. The distribution of the S2 asymmetry from the simulation and $^{137}$Cs data are overlaid in the right of Figure \ref{fig:chroma}, the data shows that the S2 asymmetry is not quite centered at 0 and has some spread. This is likely due to one PMT on the bottom which tends to see more light than the others, thus skewing the distribution. Due to these effects, reconstructing $z$ precisely is not feasible. Nonetheless the S2-asymmetry still allows for a selection of events near the axial center of the detector.
 
\begin{figure}[htbp]
    \centering
    \includegraphics[width = 0.45\textwidth]{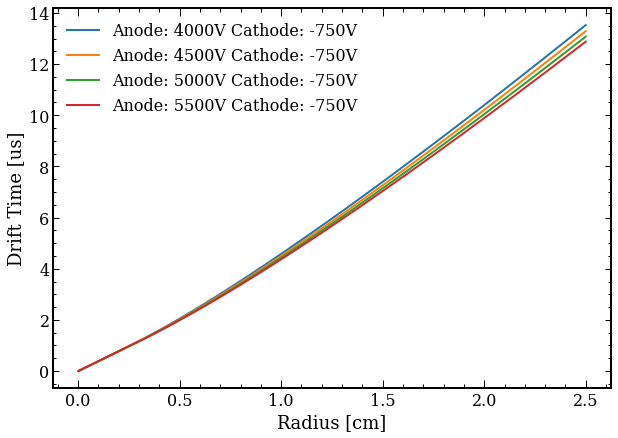}
    \includegraphics[width = 0.45\textwidth]{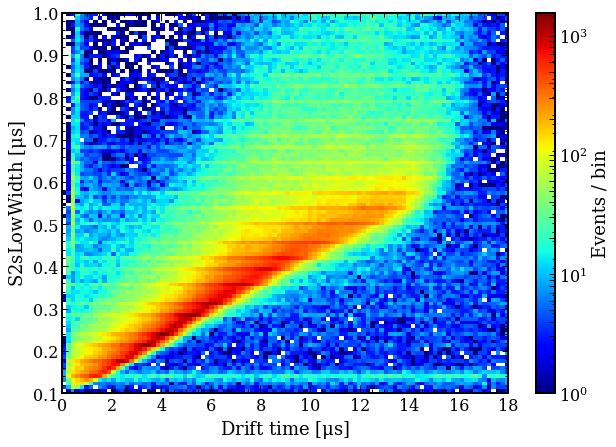}
    \caption{Left: The expected drift times as a function of event radius for an idealized detector. Right: The actual drift time vs full width at 10\% maximum distribution for a +4\,kV anode and -750\,V cathode from $^{137}$Cs.}
    \label{fig:drift_times}
\end{figure}

In order to get a good $r$ reconstruction, the field needs to be primarily in the radial direction for most of the points in the detector. However, this is not the case, as the field near the cathode wires and the top and bottom of the TPC show fringing, which gives the field lines an appreciable $\hat{\theta}$ or $\hat{z}$ component. The field effect near the top and bottom of the detector can be cut using the S2-asymmetry parameter. However, the field lines near the edge of the TPC would smear the possible values of the drift time near the cathode, as the path length no longer just depends on the $r$ position of the event, but also the $\theta$ position. Nonetheless, most events after cuts still have a drift time which is less than the maximum expected drift time. This naive expected drift time is calculated by assuming a perfectly analytic field and using the following equation:
\begin{equation}
    t_d = \int_{r_0}^{r_{anode}} \frac{dr}{v_d(E(r))}.
    \label{eq:r_dt}
\end{equation} Where $v_d(|\Vec{E}|)$ can be found using the NEST values \cite{NEST}. For an ideal RTPC, the relationship between the drift time and radius is shown in the left of Figure\,\ref{fig:drift_times} using Eq.\,\ref{eq:r_dt}. The right of Figure\,\ref{fig:drift_times} shows the S2 width as a function of drift time from $^{137}$Cs. Compared with the predicted maximum drift time of around 13$\sim$14$\mu$s in the left of Figure\,\ref{fig:drift_times}, the actual drift time is longer, possibly due to fringing field near the cathode or the uncertainty in finding the peak-center in calculating the drift time. Nevertheless, it is around 13$\mu$s of the right plot that shows a smearing of the drift times. Just as with a two-phase TPC, the events which occur further from the anode will have a larger time to diffuse compared to the events near the anode. Therefore, there is a correlation with the drift time and the S2 width as shown in Figure\,\ref{fig:drift_times} (right) from $^{137}$Cs, with most events having a width of S2 below 1\,$\mu$s.

Since this RTPC cannot resolve $\theta$ information, a cut can be done on small drift time to select for the center of the TPC where the analytic solution is accurate and Eq. \ref{eq:r_dt} can actually be used to reconstruct position. 
 
\subsection{Energy Response}
\label{subsec:energy}

\begin{figure}[htbp]
    \centering
    \includegraphics[width = 0.9\textwidth]{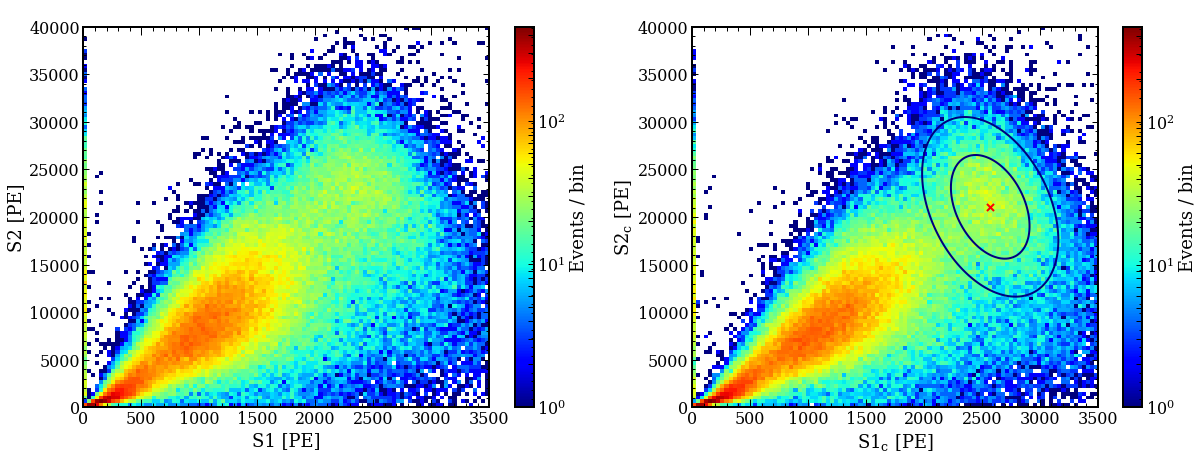}
    \caption{Left: The S1-S2 distribution for events during Cs137 runs without the field correction. Right: The same data as the left plot but with the field corrections. The drift times for these events are less than 8$\mu$s as the field here is close to the analytic field (i.e. not $\theta$ dependent). We can see that the Cs137 peak shown in the right plot is better separated from the rest of the events.}
    \label{fig:cor_uncor_cs137}
\end{figure}

The number of photons and electrons produced from an energy deposition are, on average, given by $n_\gamma = E_R L_y(E_R,\textbf{E})$ and $n_e = E_R Q_y(E_R, \textbf{E})$ respectively. Here, $L_y$ and $Q_y$ are the light and charge yields, which depend on the recoil energy $E_R$, electric field at the interaction site $\textbf{E}$, and the type of interaction. In this non-constant field, the light and charge yield will vary quite drastically from position to position, thus adding an additional smearing to the energy of an event which is far less apparent in the single phase TPC. This energy resolution gets worse with higher anode voltages as well. As such, This RTPC was only able to see the Cs137 662\,keV gamma peak clearly with a 4\,kV anode as shown in the left of Figure\,\ref{fig:cor_uncor_cs137}.

When selecting for these 662\,keV gammas, it is possible to reconcile some of the inhomogeneous field effect, since both the energy and type of interaction are known. Such a correction effectively rescales S1 and S2 to the values that they would be if the event happened at a particular reference field. The value of such a reference field is arbitrary, so this study chooses the volume averaged field for a given anode and cathode voltage. The correction is as follows:

\begin{equation}
    S1_c = S1\frac{L_y(662keV_\gamma,\langle |\textbf{E}|\rangle)}{L_y(662keV_\gamma, \textbf{E}(r(dt))} \hspace{5mm} S2_c = S2\frac{Q_y(662keV_\gamma,\langle |\textbf{E}|\rangle)}{Q_y(662keV_\gamma, \textbf{E}(r(dt))}
\end{equation} Here, the light and charge yields are from NEST, and $r(dt)$ is the radial coordinate of the interaction as a function of the drift time. Although such a correction will give a clearer energy peak than without it, this correction will only correct for the field's effect on the yields, not other potentially unforeseen physical effects, such as the electron lifetime, which is hard to extract in an inhomogeneous field.

Figure\,\ref{fig:cor_uncor_cs137} (right) gives the fit of a 662\,keV full absorption gamma peak in $(S1_c, S2_c)$ space. The result of this fit will give an $(S1_c, S2_c) = (2600, 21000)$\,PE. If we divide through by the number of expected photons and electrons (from NEST) of this type of event, then we can expect a $g_1 \approx 0.13~\rm PE/\gamma$, $g_2  \approx  0.7~\rm PE/e^-$.

\subsection{Low Energy Electronic Recoil Detection}
\label{subsec:lower}

\begin{figure}[!htbp]
    \centering
    \includegraphics[width = 0.7\textwidth]{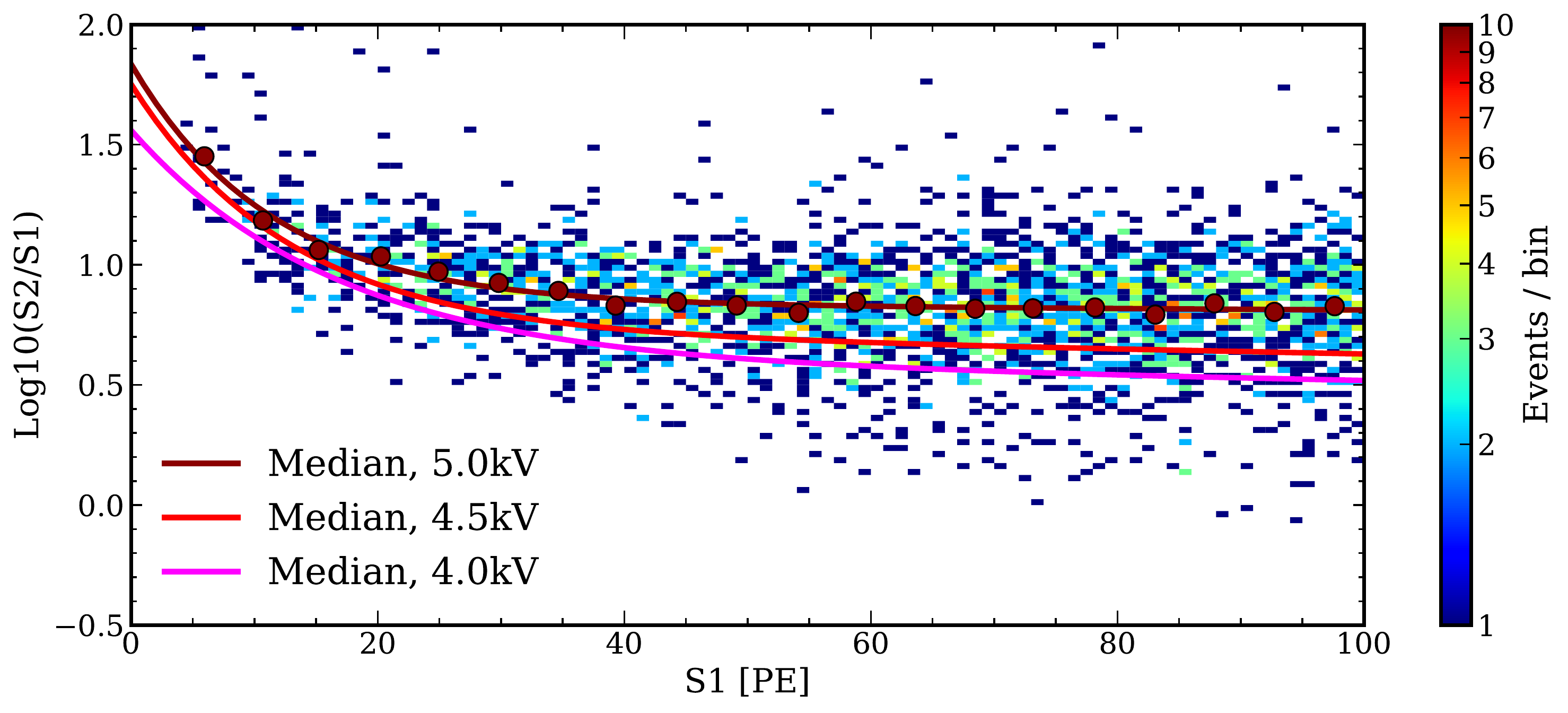}
    \caption{The electronic recoil bands for a 5.0\,kV anode voltage compared to the band medians for 4.5\,kV, and 4.0\,kV anode voltages. Cathode was set at -750\,V. We see that the median does increase as the anode voltages increase. However, S2/S1 never gets to around 100 as it does in a two phase detector, thus making electrons hard to detect.}
    \label{fig:bands}
\end{figure}

Despite the overall low S2 amplification, we can still get the idea that the ionization signal is amplified as the anode voltages increases. This is clear when looking at the low energy electronic recoil bands shown in Figure \ref{fig:bands}.

In order to get a good amplification for S2 signals, the anode voltage needs to be very high. However, increasing the anode voltage comes at the cost of an increased rate of light emissions, thereby making it hard to identify small S1s. Furthermore, the low $g2$ value in the RTPC means that single electrons are on the order of a few PE in size, and the short electroluminescence region means that they only have several nanoseconds of drift time. So a pileup of these light emissions can also affect the identification of the actual S2 signal from an event as well.

\begin{figure}[htbp]
    \centering
    \includegraphics[width = 0.49\textwidth]{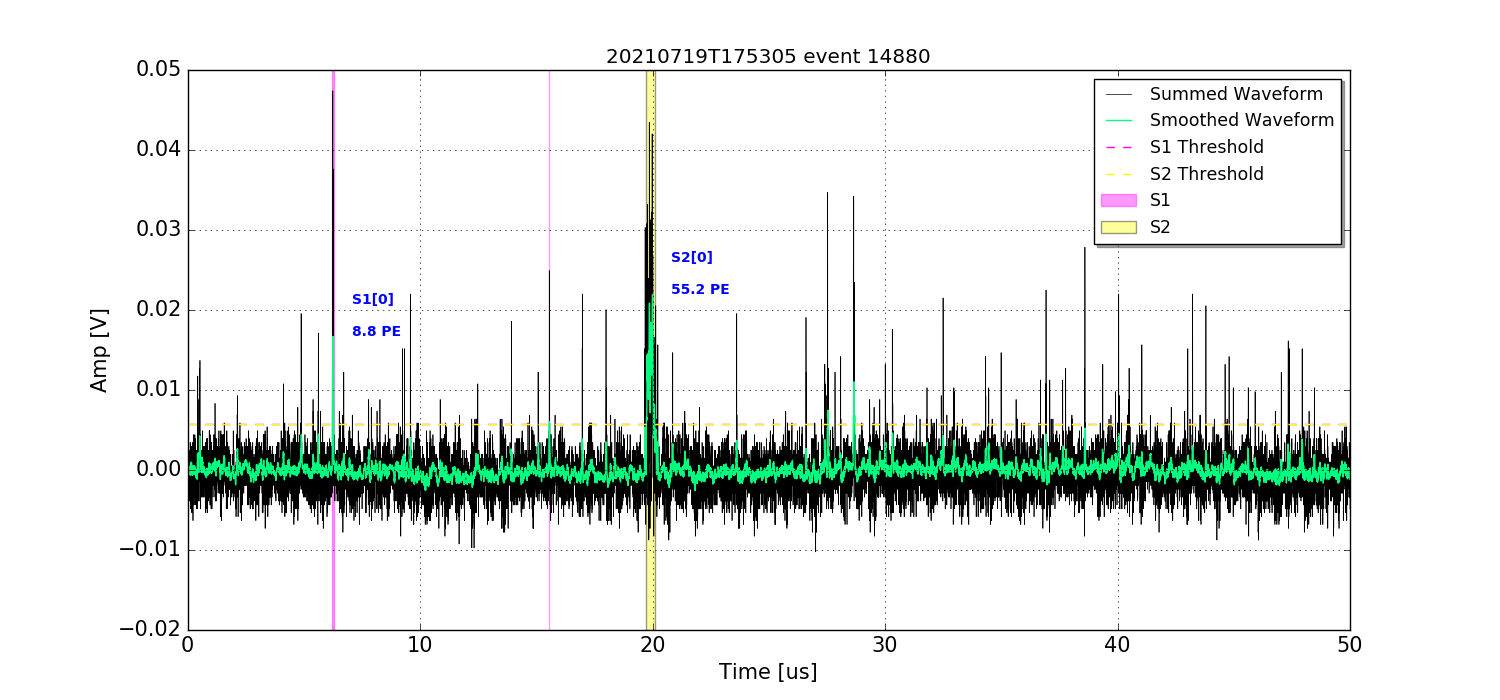} 
    \includegraphics[width = 0.49\textwidth]{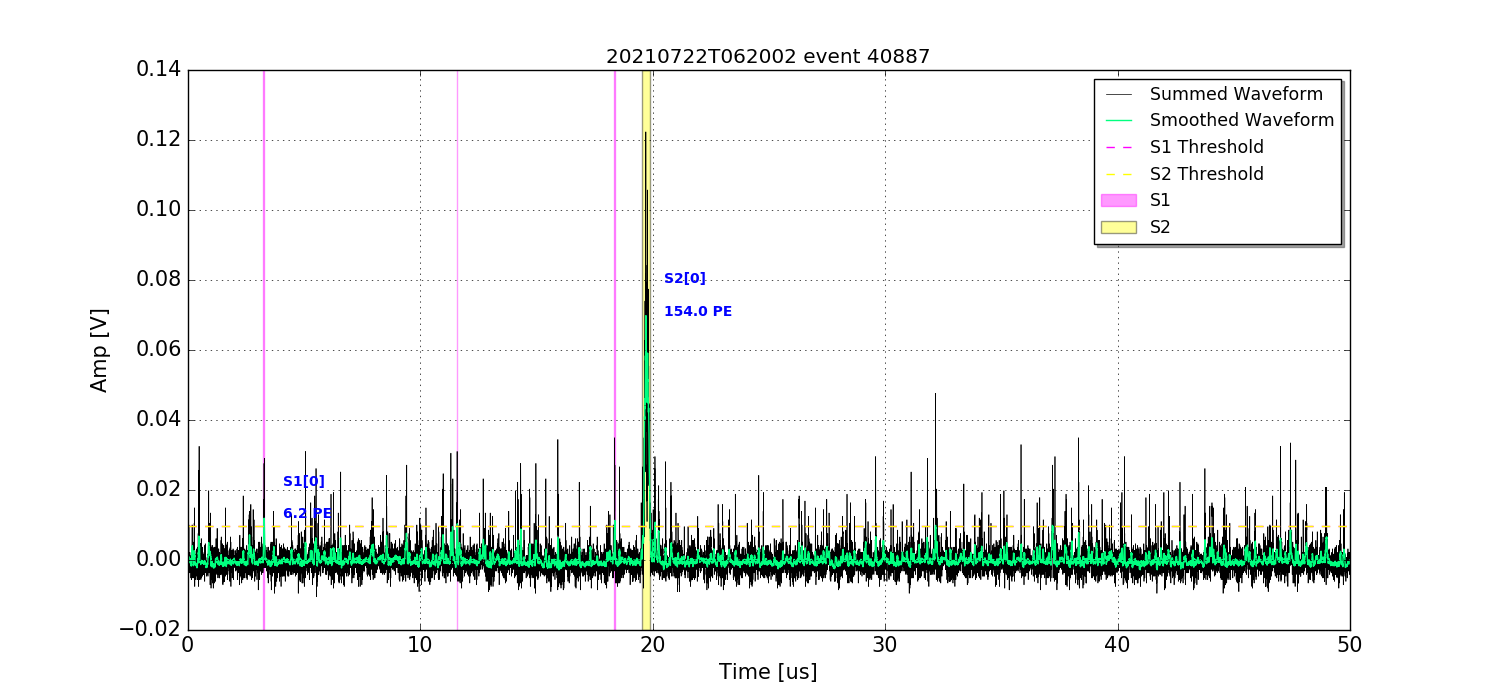} 
    \includegraphics[width = 0.49\textwidth]{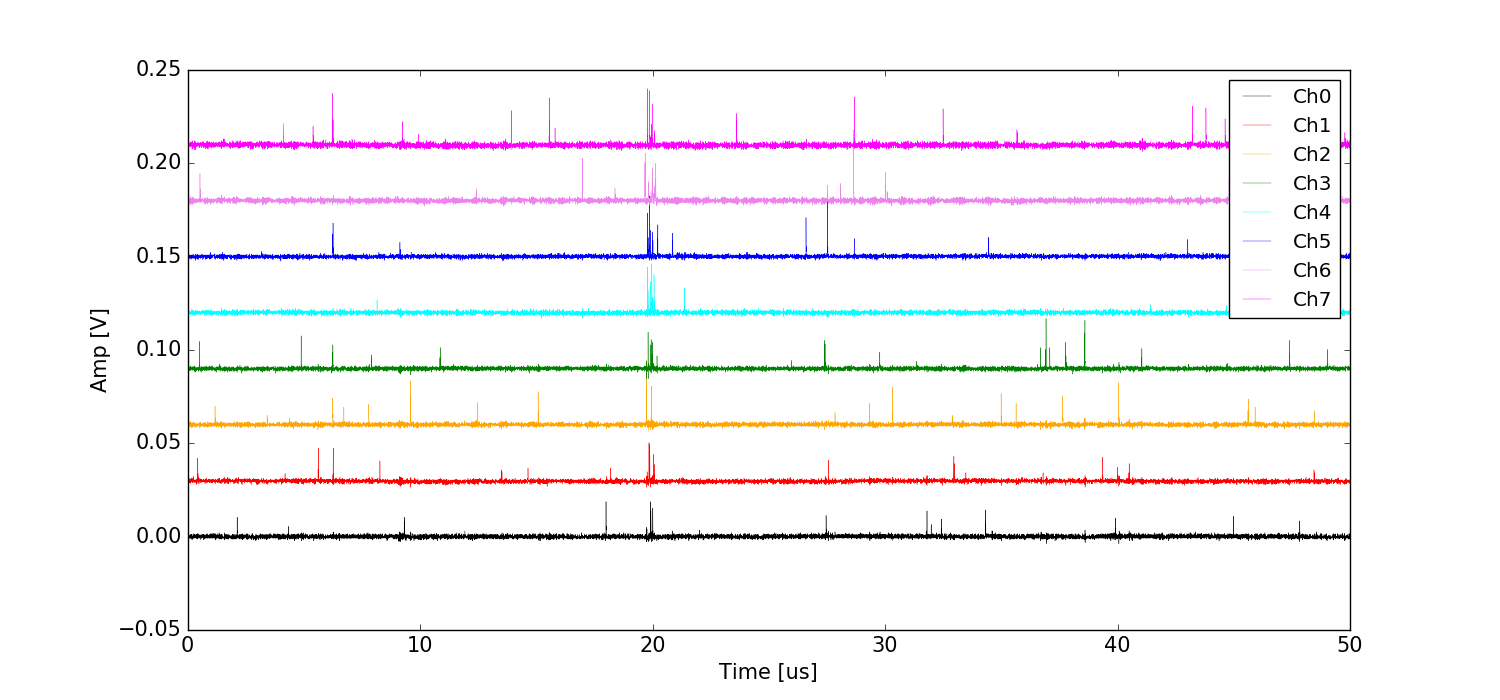}
    \includegraphics[width = 0.49\textwidth]{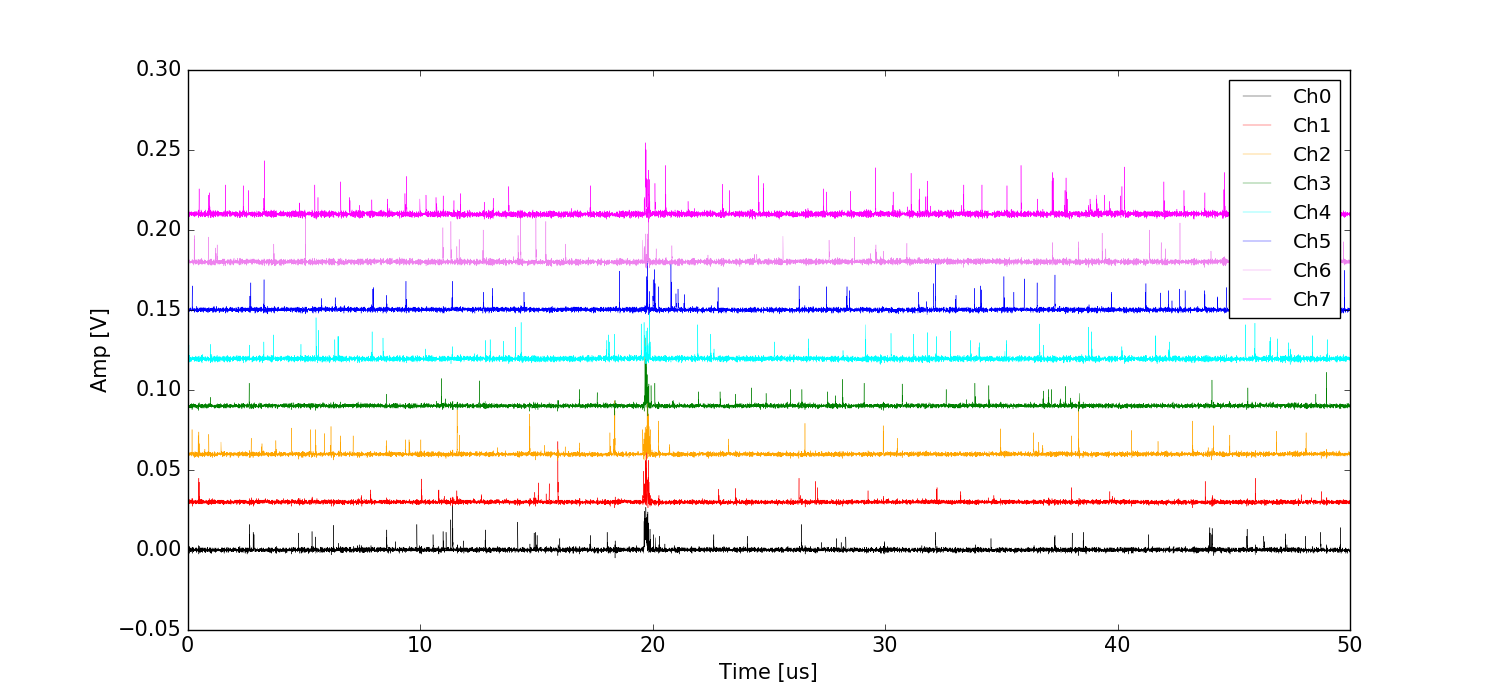}
    \caption{Left (top): Summed waveform of a low energy event at 4\,kV on the anode wire. Left (bottom): individual PMT waveforms of the Left (top) event. Right (top): Summed waveform of a low energy event at 5 kV on the anode wire. Right (bottom): individual PMT waveforms of the Right (top) event.}
    \label{fig:lightemission}
\end{figure}

Looking at left of Figure \ref{fig:lightemission}, the smoothed waveform (green) and the per-PMT waveform of the top event from 4\,kV anode shows that although the identified S1 does seem to have noisy contenders, five different PMTs see light at around the time the identified S1 happens, and the waveform is not filtered out by smoothing. However, the event at 5kV anode shows in the right of Figure \ref{fig:lightemission} does not show such a clear identification. The per-PMT waveforms are clearly noisier, with much more single PE level signals spread throughout each PMT waveform. This makes it difficult to resolve both small S1s and single to few-electron S2s.

\section{Limitations of the Liquid Xenon RTPC for Potential Applications}
\label{sec:potential}

With its high stopping power and abundant scintillation and ionization yield, LXe is a very good target medium for radiation detection. The RTPC simplified the design, construction and operation of the LXeTPC compared to a conventional two-phase TPC used for dark matter searches~\cite{Akerib:2016vxi,Cui:2017nnn,Aprile:2017aty}. Using only the photo-sensors to detect both the primary scintillation and ionization converted electroluminescence light removes the need of the charge-readout electronics such as used in a LXe gamma ray imaging telescope~\cite{Aprile:2008ft}. In this study, we show that MeV gamma rays as well as electronic recoils down to about 10 keV can be detected with such a detector design. However, such a simplified design reduces the energy and position resolution. This technique might be useful for limited applications for counting low energy gamma rays where high energy resolution or position resolution is not required. 

The small amplification factor for the ionization signal with only about 1 PE/e- was detected in our experiment. Increasing the electric field around the anode wire will increase the amplification factor, but we observed increasing light emissions that would limit the detectability of single photons or single electrons from the true physical events. This light emission might be associated intrinsically with the strong electric field around the anode wire which can't be suppressed thus making such a design not feasible for ultra-low energy event detection such as the search for light dark matter or the detection of CEvNS from reactor neutrinos.

\acknowledgments

We thank A. Kopec and R. Lang for their help on the PMT bases. This research is sponsored by the US Defense Advanced Research Projects Agency under grant number HR00112010009, the content of the information does not necessarily reflect the position or the policy of the Government, and no official endorsement should be inferred.

\bibliographystyle{unsrt}
\bibliography{bibliography}

\end{document}